\documentclass{jaa}

\usepackage{graphicx}
\usepackage{aas_macros}
\usepackage{natbib}
\begin{document}

\title{The Nature of Coherent Radio Emission from Pulsars}


\author{Dipanjan Mitra\textsuperscript{1}\textsuperscript{2}\textsuperscript{3}}
\affilOne{\textsuperscript{1,}National Center for Radio Astrophysics, TIFR, Pune 411007, India.\\}
\affilTwo{\textsuperscript{2,}Physics Department, University of Vermont, USA, Burlington VT 05405.\\}
\affilThree{\textsuperscript{3}Janusz Gil Institute of Astronomy, University of
Zielona G\'{o}ra, ul. Szafrana 2, 65-516 Zielona G\'{o}ra, Poland\\}


\twocolumn[{

\maketitle

\corres{dmitra@ncra.tifr.res.in}

\msinfo{14 April 2017}{}{21 June 2017}

\begin{abstract}
The pulsar radio emission originates from regions below 10\% of the light 
cylinder radius. This requires a mechanism where coherent emission is excited
in relativistic pair plasma with frequency $\nu_{cr}$ which is below the plasma
frequency $\nu_{\circ}$ i.e. $\nu_{cr} < \nu_{\circ}$. A possible model for the
emission mechanism is charged bunches (charged solitons) moving 
relativistically along the curved open dipolar magnetic field lines capable of 
exciting coherent curvature radio emission. In this article 
we review the results from 
high quality observations in conjunction with theoretical models to unravel 
the nature of coherent curvature radio emission in pulsars. 
\end{abstract}


}]


\doinum{12.3456/s78910-011-012-3}
\artcitid{\#\#\#\#}
\volnum{123}
\year{2017}
\pgrange{1--15}
\setcounter{page}{1}
\lp{15}

\section{Introduction}
A pulsar is a fast spinning, highly magnetized neutron star, capable of 
generating beamed radiation observed as periodic pulses observed approximately 
over the entire electromagnetic spectra. 
\citet{1969ApJ...157..869G} pointed out that the neutron star can generate 
enormous co-rotational electric field around the star and charges can be pulled
out of the star and create a charge separated magnetosphere with density 
$n_{GJ}= \mathbf{\Omega.B}/2\pi e c$, (where $\Omega = 2\pi/P$, $P$ being 
the rotation period of the star and $\mathbf{B}$ its magnetic field), and the
magnetosphere can be divided into open and closed field line regions, with a 
relativistic flow of charges along the open field lines leading to the observed
emission. This seminal idea went through several refinements and presently it 
is understood that an additional source of plasma is essential to establish the
relativistic flow of charges along open dipolar field lines (see for e.g. 
\citealt{1999PhR...318..227M}, \citealt{2011ASSP...21..139S}, 
\citealt{2016JPlPh..82e6302P}). \citet{1970Natur.227..465S,1971ApJ...164..529S}
was amongst the first to suggest the magnetic pair production by $\gamma$-ray 
photons ($>$ 1.02 Mev) in strong magnetic ($\sim 10^{12}$ G) as a source of 
plasma. 

Pulsars are known to slow down by expending large amount of energies as 
magnetic dipole radiation as well as particle winds and electromagnetic 
radiation (slow down energy $\dot{E} \sim 10^{30-38}$ ergs~s$^{-1}$).
The majority of the electromagnetic emission is in the form of X-ray and 
$\gamma$-rays with only a tiny fraction ($\sim10^{28}$ ergs~s$^{-1}$) emitted 
in the radio wavelengths, which when converted to brightness temperature yields
extremely high values of $\sim$10$^{28-30}$K. Only a collective or coherent 
mechanism, either by charged bunches (e.g. \citealt{1975ApJ...196...51R}, RS75 
hereafter) or a maser mechanism that arises due to growth of plasma 
instabilities (e.g. \citealt{1991MNRAS.253..377K}), can excite the coherent 
radio emission. 

This continues to be a challenging problem in astrophysics (see for e.g. \citealt{1995JApA...16..137M}). However, in recent 
years significant progress has been made thanks to high quality observations as
well as enhanced theoretical developments. In this article we show how various 
observations tend to favour the idea that the coherent radio emission in 
pulsars are excited by curvature radiation from charged bunches.  

\section{Observational Constraints on pulsar radio emission}
Radio pulsars 
exhibit a wide period range from $\sim$1.3 milliseconds to 8.5 seconds. 
Around periods of 30 milliseconds the pulsar population separates into two groups, the  
millisecond pulsars ($P <$ 30 msec) and the normal pulsars ($P >$ 30 msec), 
and the latter are the focus of discussion in this article. 
The pulsed emission is usually restricted to a window called the main pulse 
(MP), the width of the window depends on observer's line of sight geometry 
across the emission beam. In certain specialized geometries an inter-pulse (IP) 
emission, located 180$^{\circ}$ away from the MP, is also observed. In very 
rare cases an additional pre/post-cursor (PC) emission component is seen 
connected to the MP by a low level bridge emission. More recently a continuum 
off-pulse (OP) emission has also been detected in some long period pulsars. 
The single pulses are highly variable which can modulate in time 
although averaging a few thousand pulses produce a stable full stokes 
pulse profile which is a signature of the particular pulsar. 
In this section we summarize the observations, both single pulses as well as 
average profiles, whose interpretation only assumes beamed radiation by 
relativistic flow of charges along magnetic open dipolar field lines.

\subsection{Average profile and Geometry} 
The single pulses corresponding to the MP are structured and consists of one
or more Gaussian like subpulses. In average profiles these subpulses form 
distinct components at specific locations. The centrally located component is 
called ``core'' which is surrounded by concentric pairs of ``cones'' 
(\citealt{1976ApJ...209..895B}, \citealt{1983ApJ...274..333R}). 

Pulsar emission is highly linearly polarized and the corresponding polarization
position angle (PPA) across the pulsar profile shows a characteristic S-shaped 
swing. This has been interpreted using the rotating vector model (RVM, 
\citealt{1969ApL.....3..225R}), as a signature of emission arising from open 
dipolar magnetic field lines pulsar associated with the line of sight geometry,
\begin{equation}
\begin{aligned}
\scriptsize
\Psi = \Psi_{\circ} + \\ 
\tan^{-1}\left(\frac{\sin(\alpha)\sin(\phi-\phi_{\circ})}
{\sin(\alpha + \beta) \cos(\alpha) - \sin(\alpha) \cos(\alpha + \beta)\cos(\phi-\phi_{\circ})}\right) \\ 
\end{aligned}
\label{eq1}
\end{equation}
where $\alpha$ is the angle between the rotation axis and the dipolar magnetic 
axis and $\beta$ is the angle between the magnetic axis and the observers line 
of sight. The point of steepest gradient (SG) of the PPA traverse lies in the 
fiducial plane containing the rotation and magnetic axis, and the slope of the 
PPA at SG is $R_{ppa}=\mid d\Psi/d\phi\mid=\sin(\alpha)/\sin(\beta)$.
Here $\phi_{\circ}$ is the longitude corresponding to SG with the PPA give as 
$\Psi_{\circ}$. The PPA traverse is often complicated by the presence of 
orthogonal polarization modes (OPM), and single pulse studies are needed to 
unravel the underlying RVM. It should be noted that using eq.(\ref{eq1}) to 
obtain the geometrical parameters, particularly $\alpha$ and $\beta$, is futile
as they are highly correlated and no meaningful constraint can be derived 
(\citealt{1997A&A...324..981V}, \citealt{2001ApJ...553..341E}). However, 
$\Psi_{\circ}$ and $\phi_{\circ}$ are better estimated from this fit. The RVM 
is valid for any diverging set of magnetic field lines (e.g. in off-centered 
dipole as seen in \citealt{2017MNRAS.466L..73P}) and with the commonly used model of
a star centered global magnetic dipole is being merely a good assumption. 

The locus of the open dipolar magnetic field line in the inner magnetosphere 
is roughly circular (e.g. \citealt{2004ApJ...614..869D}).
Identifying the leading and trailing edge of the profile with the last open 
field lines arising from same emission height, the half opening angle or beam radius $\rho^{\nu}$ can be computed using spherical trigonometry as (Gil 1981),
\begin{equation}
\sin^2(\rho^{\nu}/2)=\sin(\alpha+\beta) \sin(\alpha) \sin^2(W^{\nu}/4)
+\sin^2(\beta/2)
\label{eq2}
\end{equation}
where $W^{\nu}$ is the width of the profile at frequency $\nu$. In general 
$W^{\nu}$ decreases with increasing frequency (known as radius-to-frequency 
mapping, RFM) and hence $\rho^{\nu}$ is also a function of frequency 
(see \citealt{2002ApJ...577..322M}). 
For emission arising from last open dipolar field line 
at a height $h_{em}^{\nu}$ from the center of the star, $\rho^{\nu}$ can be 
related to $h_{em}^{\nu}$ as 
$\rho^{\nu} = 85^{\circ}.9 \sqrt{ 2 \pi h_{em}^{\nu}/cP}$, where 
$c$ is the velocity of light. 
For a neutron star of radius $h_{em} = 10$ km the full opening angle 
at the polar cap is given by $2\rho_{pc} = 2.45^{\circ}~P^{-0.5}$.
Taking the ratio of $\rho^{\nu}$ with $\rho_{pc}$  for a pulsar with
period $P = 1$ sec the emission radius $h_{em}^{\nu}$ can be written as, 
\begin{equation}
h_{em}^{\nu} = 10 P (\rho^{\nu}/1.23^{\circ})^2~~~\rm{km.}
\label{eq3}
\end{equation}

We explore the implications of the period dependence of opening angle 
($\rho^{\nu}\propto P^{-0.5}$). \citet{1990ApJ...352..247R,1993ApJ...405..285R} 
estimated $\rho^{\nu}$ using the half-power widths of the core and conal 
components at 1 GHz, and demonstrated that $2\rho^{1GHz}_{core} = 
2.45^{\circ}~P^{-0.5}$, $\rho^{1GHz}_{in} = 4.3^{\circ}~P^{-0.5}$ and 
$\rho^{1GHz}_{out} = 5.7 P^{-0.5}$. We summarize the arguments that led to 
these results. \citet{1990ApJ...352..247R} noticed that when the half-power 
width of the core component was plotted with $P$, a lower boundary line (LBL) 
$2.45^{\circ} P^{-0.5}$ existed. Several IPs were found to have core components (hence $\alpha \sim 90^{\circ}$ and $\beta \sim 0^{\circ}$) with widths along
the LBL. \citet{1990ApJ...352..247R} suggested that the $W^{1GHz}_{core}$ above
the LBL were due to non-orthogonal ($\alpha \neq 90^{\circ}$, $\beta 
\sim 0^{\circ}$) geometry. 
Thus using eq(\ref{eq2}), (\ref{eq3}) and 
small angle approximation for $\rho^{1GHz}_{core}$ and $W^{1GHz}_{core}$
a connection was established as,
\begin{equation}
2\rho^{1GHz}_{core} = W^{1GHz}_{core} = 
2.45^{\circ} P^{-0.5}/\sin(\alpha)
\label{eq4}
\end{equation}
This scheme found $\alpha$ for a pulsar with core emission and consequently 
using the $R_{ppa}$ one could estimate $\beta$. 
\citet{1993ApJ...405..285R,1993ApJS...85..145R} used $\alpha$ and $\beta$ 
obtained from core measurements and using eq.(\ref{eq2}) calculated 
$\rho^{1GHz}_{in}$ and $\rho^{1GHz}_{out}$. Thus, the estimation of 
$\rho^{1GHz}_{in,out}$ using core widths automatically transfers the $P^{-0.5}$
dependence. Rankin argued that the LBL for core emission can be explained by 
approximating the core component as a bi-variate Gaussian with the emission 
arising from the entire surface of the polar cap. In order to recover the 
$2\rho^{1GHz}_{core} = 2.45^{\circ} P^{-0.5}$ dependence, the half-power points 
should correspond to the last open dipolar field lines.
While this argument is compelling, physically it is difficult to conceive the 
coherent radio emission being generated near the polar cap.

\begin{figure}[!t]
\includegraphics[width=.75\columnwidth,angle=-90]{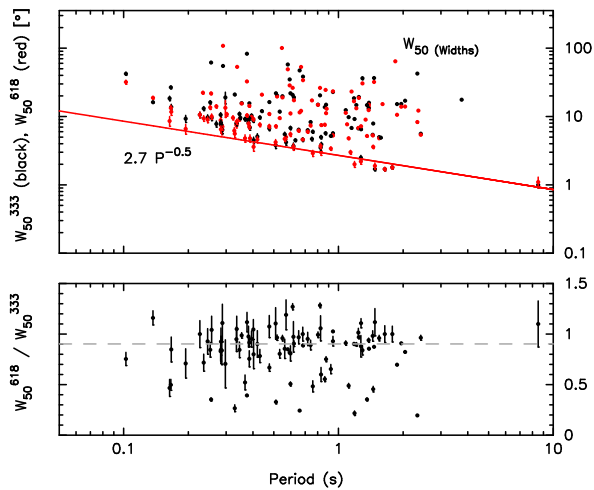}
\caption{The top panel shows the distribution of widths (measured at 
outer half-power points) with pulsar period for MSPES sample at 333 and 618 MHz.
The points near the LBL consist of both core and cone pulsars and 
is represented by a guiding line $2.7 P^{-0.5}$. The bottom panel
show the ratio of widths between 618 and 333 MHz, and a mean value of
0.8 is due to RFM. This figure is reproduced from Fig 2. of \citet{2016ApJ...833...28M}.  
}\label{fig1}
\end{figure}

In recent works \citet{2011MNRAS.417.1444M,2012MNRAS.424.1762M} showed that the
distribution of half-power width of a large number of pulsars
with period could reproduce the LBL. In their sample no distinction was made 
between any profile class and hence the LBL existed for both core and conal 
pulsars and were dominated by the lowest angular structures in the average 
profile. They argued that the LBL is consistent with core and cone emission 
arising from about 50 stellar radii. The numerical factor $2.45^{\circ}$ was
related to the smaller structures in the polar cap 
and the $P^{-0.5}$ dependence followed from the dipolar nature of the 
open field lines. As we will discuss below, there is observational evidence 
that the core and conal emission arises from similar heights of a few hundred 
kilometers. The finite emission height of core's would imply that the assumption 
that core emission arises from last open field lines is invalid.

The presence of the LBL is also seen in the Meter-wavelength Single-Pulse 
Polametric Emission Survey (MSPES) at 333 and 618 MHz 
(\citealt{2016ApJ...833...28M}), reproduced in Fig.(\ref{fig1}). A more 
detailed study of this data set have revealed that core and cone separately 
follow the $P^{-0.5}$ scaling relation (Skrzypczak et al. 2017~in preparation).
These studies also show that $P^{-0.5}$ scaling is a natural consequence of the
dipolar fields only if $\rho^{\nu}$ corresponds to the last open field lines. 
The components that arise from a certain fixed height and occupy inner regions
of the open magnetic field lines do not scale as $P^{-0.5}$. In this case 
the observed $P^{-0.5}$ dependence must have a different physical origin.

\begin{figure}[!t]
\includegraphics[width=.9\columnwidth,height=2.0\columnwidth,angle=-0]{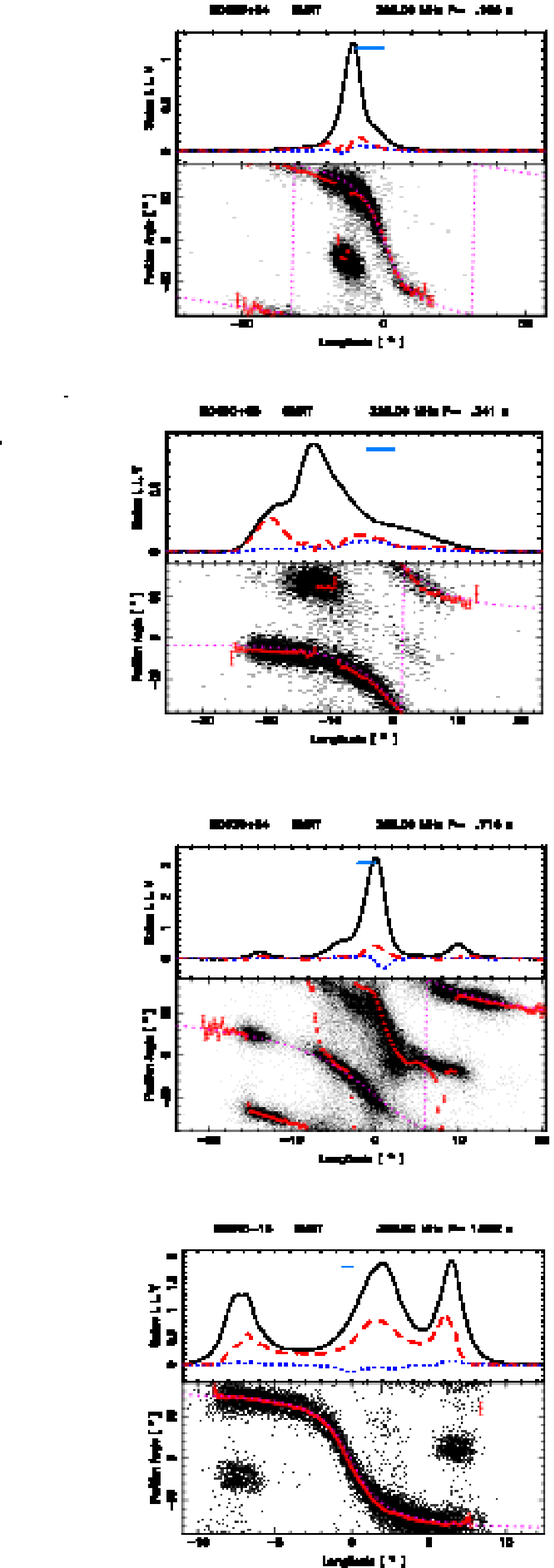}
\caption{These plots show the application of delay-radius method of finding 
emission heights given by eq.(\ref{eq6}). The plots from top to bottom 
corresponds to pulsar with increasing period. In each plot the top panel shows 
the average stokes parameters and the bottom panel shows the polarization 
histograms. The zero longitude in each plot correspond to the SG point of the 
PPA and the blue horizontal bar on the top panel shows the delay $\Delta \phi$ 
The top two plots uses data that was published in Fig.(1) and Fig.(A1) 
of \citet{2011ApJ...727...92M} and the bottom plot from 
Fig.(1) of \citet{2014ApJ...794..105M}.}
\label{fig2}
\end{figure}
 
\subsection{Pulse shape and phenomenology}
The parameters of the emission beam as well as the line of sight geometry, as 
discussed above, forms the the basis of the `core-cone' model of emission beam.
The pulsar profiles can be broadly classified into the following categories.
The central line of sight profiles with three and five components are known as 
triple ({\bf T}) amd multiple ({\bf M}) respectively. More tangential 
line of sight geometries with one or two components are called conal single 
({\bf S$_d$}) or conal double {\bf D} profiles.
More detailed discussion on the profile morphology is carried out in  
\citet{1976ApJ...209..895B, 1983ApJ...274..333R,1990ApJ...352..247R,1993ApJ...405..285R,1993A&A...272..268G,1997MNRAS.288..631K}.
These studies reveal the shape of the pulsar radio emission beam, which can be 
thought of as an emission pattern projected in the sky, to be composed of a 
central core emission surrounded by two nested inner and outer cones. 
\citet{1999A&A...346..906M} carried out a multi-frequency analysis and found 
three nested cones with opening angle given by $\rho^{\nu} = 4.8 K (1 + 66 
\nu^{-1}) P^{-0.5}$ where $K=0.8,1,1.2$ for the three cones respectively. They 
also estimated the angular width of each conal ring to be about 20\% of 
$\rho^{\nu}$. 
There are studies with contradictory viewpoint about the shape of pulsar 
emission beam. For example \citet{1988MNRAS.234..477L,2001MNRAS.320L..35H} 
considers the pulsar beam to be composed of random patches with the pulse shape
independent of the line of sight geometry. Their conclusions were supported by 
``partial-cone'' pulsars, where the SG point of the PPA was seen at one edge of 
the profile, giving the impression that part of the emission from the beam was 
missing. \citet{2011ApJ...727...92M} carried out a detailed 
single pulse analysis and showed the SG point to lie on the trailing edge of 
the profile in the ``partial-cone'' pulsars (see Fig.\ref{fig2}).
This is indicative of relativistic beaming effect and particularly the presence
of single pulse flaring property established the pulse profile shapes to be 
consistent with the core-cone model.
The nested core-cone structure is based on only the MP emission and
\citet{2015ApJ...798..105B} showed that the PC components 
seen in a small sample of pulsars could not be reconciled with the core-cone 
picture and they likely have different locations within the magnetosphere.

\begin{figure*}
\centering\includegraphics[height=.20\textheight,angle=0]{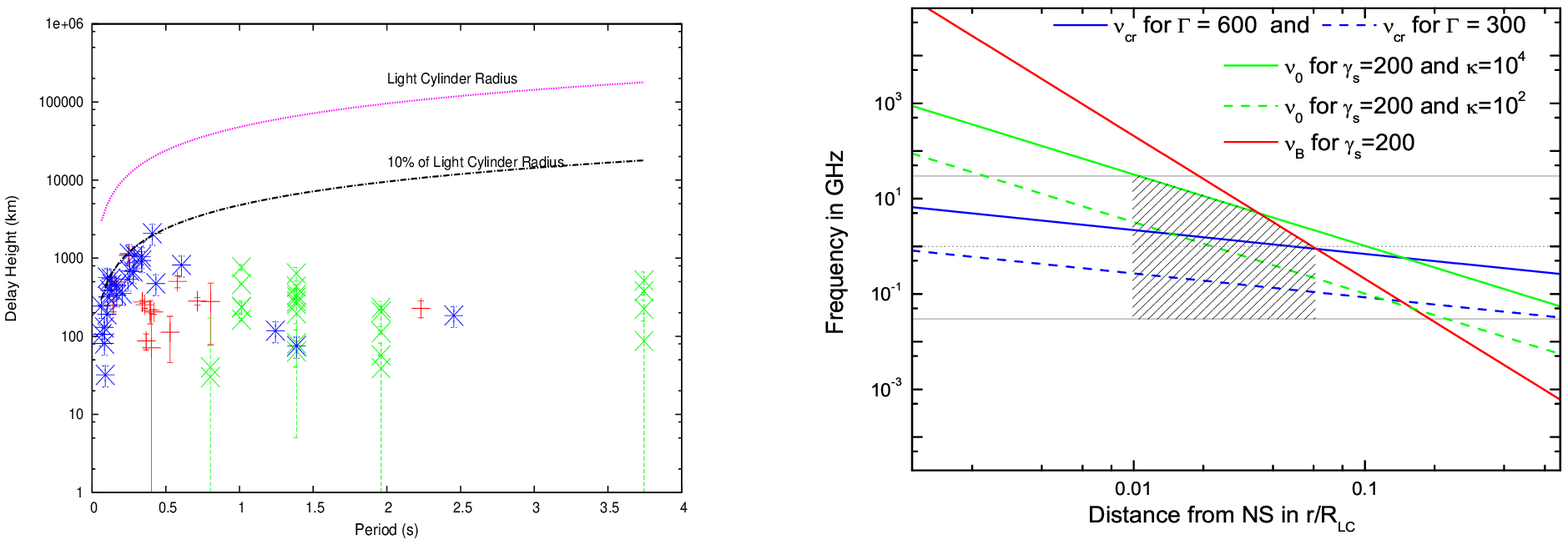}
\caption{The left panel shows period versus delay-radius emission heights estimated using 
eq(\ref{eq6}) obtained from various sources only for cases where $\Delta \phi$ is positive. 
The blue points are taken from Table (3) of \citet{2008MNRAS.391.1210W}, the green 
points are obtained from table (3) of \citet{2004A&A...421..215M} and red points 
are obtained from table (A4) of \citet{2011ApJ...727...92M}.  
The right panel plot shows the plasma parameters as a function of the fraction of light
cylinder which is reproduced from Fig.(4) of \citet[, see text for details]{2014ApJ...794..105M}.}
\label{fig3}
\end{figure*}

\subsection{Emission Heights} 
The radio emission height in pulsars is estimated using three different 
techniques namely the geometrical method, delay-radius method and 
scintillation studies. 
The geometrical method gives an estimate of emission height based on 
eq.(\ref{eq3}), which has unresolved issues as discussed in the previous
subsection.
\citet{1990ApJ...352..247R,1993ApJ...405..285R,1993ApJS...85..145R} estimated 
$h_{em, in,out}^{1GHz}$ to be around 130 and 220 km respectively which were 
largely independent of the pulsar period. Using a similar approach, but extending
the widths to significantly lower thresholds over multiple frequencies, 
\citet{1997MNRAS.288..631K} found the height of the outer conal emission to 
depend mildly on pulsar age and period: $h_{em} \approx 550 \nu^{-0.21} 
\tau_6^{-0.1}P^{0.33}$ km, $\tau_6$ is characteristic age in $10^6$ yr s. 

The delay-radius method was suggested by \citet[,hereafter BCW]{1991ApJ...370..643B}. The method utilizes the fact that the emitting plasma in the co-rotating 
frame has slightly bent trajectories in the direction of rotation in the
observer's frame.
BCW showed that if the emission across the profile originates at a fixed height
$h_{em} << R_{lc}$ (where $R_{lc} = cP/2\pi$ is the light cylinder radius)
then the shape of the PPA traverse in eq.(\ref{eq1}) is modified such that the 
phase is shifted by $\Delta \phi \approx 4h_{em}/R_{lc}$ with respect to the center 
of the pulse profile and the PPA undergoes a downward shift 
$\Delta\Psi=-(10/3)(h_{em}/R_{lc})\cos(\alpha)$. \citet{2001ApJ...546..382H} (also 
see \citealt{2012ApJ...746..157K,2012ApJ...754...55K,2013ApJ...769..104K}), 
showed that due to polar current $J$ the PPA undergoes a vertical upward shift 
$\Delta \Psi = (10/3)(h_{em}/R_{lc})J/J_{GJ} \cos(\alpha)$, $J_{GJ}$ is the 
Goldreich-Julian current density. When $J=J_{GJ}$, $\Delta \Psi$ exactly 
cancels the shift due to the delay-radius relation. 
\citet{2008MNRAS.391..859D} provided a lucid physical description of this 
effect and derived a modified form of eq.(\ref{eq1}) in the observer's frame 
as:
\begin{equation}
\begin{aligned}
\Psi =  \Psi_{\circ} + \\
         \tan^{-1}\left(\frac{\sin(\alpha)\sin(\phi-\phi_{f}-2h_{em}/c)}
{\sin(\alpha + \beta) \cos(\alpha) - \sin(\alpha) \cos(\alpha + \beta)\cos(\phi-\phi_f - 2h_{em}/c)}\right) 
\end{aligned}
\label{eq5}
\end{equation}
Here $\phi_f$ is fiducial pulse phase corresponding to the plane containing the
rotation and magnetic axis.
If the emission altitude varies significantly across the pulse profile then 
the PPA traverse can be distorted. However, if emission arises from the same 
height, $\phi_f = (\phi_c + \phi_{PPA})/2$, where $\phi_c$ is at the center of 
the pulse profile and $\phi_{PPA}$ is the phase at the SG point. The emission 
height $h_{em}^{delay}$ is given as 
\begin{equation}
h_{em}^{delay} = \frac{c}{4} \frac{\Delta\phi}{360} P~~~{\rm km,}
\label{eq6}
\end{equation}
where $\Delta \phi = (\phi_{PPA}-\phi_c)$. Estimating the emission height using
eq.(\ref{eq6}) involves two steps, first is fitting eq.(\ref{eq1}) to the PPA traverse
to establish a good model for the RVM and obtain an estimate of $\phi_{PPA}$; 
and secondly to find the midpoint of the profile $\phi_c$ by measuring the 
phase $\phi_l$ and $\phi_t$ at the leading and trailing edge of the profile 
such that $\phi_c = \phi_l +(\phi_t-\phi_l)/2$. The delay-radius method has 
been used to estimate the emission heights in a large number of pulsars and 
is around $h_{em}^{delay} \sim 200-500$ km (BCW, \citealt{1997A&A...324..981V,2004A&A...421..215M,2011ApJ...727...92M,2008MNRAS.391.1210W}). 
\citet{2004A&A...421..215M} notes that there are many factors that affect 
estimates of $h_{em}$ with upto 30\% systematic errors.
For example, $\phi_l$ and $\phi_t$ are usually measured at the half-power or 
10\% level of the profile which might not correspond to the last open field 
line, or/and the assumption that overall emission arising from the same height 
might not be correct. In fact for some individual cases $\Delta \phi$ can even 
be negative, and thus delay-radius height estimates should be used in a 
statistical sense. It is important to note that eq.(\ref{eq6}) has been derived
assuming a linear theory with first order terms of $h_{em}/R_{lc}$ retained. 
\citet{2008MNRAS.391..859D} pointed out that the systematic error due to this 
approximation is about $(h_{em}/R_{lc})^{0.5}$ and hence for $h_{em}/R_{lc} 
\sim 0.1$, the error is about 30\%. \citet{2008MNRAS.391..859D} also found the 
theory to be valid for $h_{em}^{delay}/R_{lc} < 0.1$ or $\Delta\phi < 
20^{\circ}$, since second order effects like magnetic field sweep-back, polar 
currents or Shapiro delay becomes important. Fig.(\ref{fig2}) shows the 
application of the delay-radius method in three pulsars with different periods. 
In Fig.(\ref{fig3}) left panel we show estimates of $h_{em}^{delay}$ from 
multiple sources (see figure caption) and plot them as a
function of period. The heights are roughly constant with period and they lie 
below
10\% of $R_{lc}$. The delay-radius method has also been used to identify the 
location of the core emission. For example, in the core-dominated pulsar PSR 
B1933+16 \citet{2016MNRAS.460.3063M} both the core and conal emission to
be significantly delayed with respect to $\phi_{PPA}$ were shown, thus suggesting similar 
emission heights for the overall emission. There are a few studies that have
used the center of the core emission as the fiducial point 
(\citealt{1998ARep...42..388M,2001ApJ...555...31G,2003ApJ...584..418G,
2009MNRAS.393.1617K}) with the delay-radius relation measuring the conal 
emission heights with respect to the core component. These studies have 
suggested that the core emission is emitted slightly lower than the conal 
emission. 

The third method uses the fact that the emission from compact emission region 
of pulsars traverse through the interstellar medium which act as a varying 
lens modulating the pulsar signal (\citealt{1983ApJ...268..370C}). The extent 
of modulation depends on the transverse size of the source, and can be 
estimated using high spatial resolution interferometry. This method has been 
applied on a small number of pulsars, with accurate results available for the 
Vela pulsar. The estimated spatial extent of the emission source in Vela pulsar
is about 4 km with the corresponding radio emission altitude of about 340 km 
(see e.g. \citealt{2012ApJ...758....8J}). 

In summary three independent methods finds the pulsar radio 
emission altitude $h_{em} \sim 200-500$ km from the neutron star surface. This
is a crucial input into theoretical models of pulsar radio emission mechanism.

\subsection{Evidence of Curvature radiation}
\begin{figure*}
\centering\includegraphics[height=.21\textheight,angle=-0]{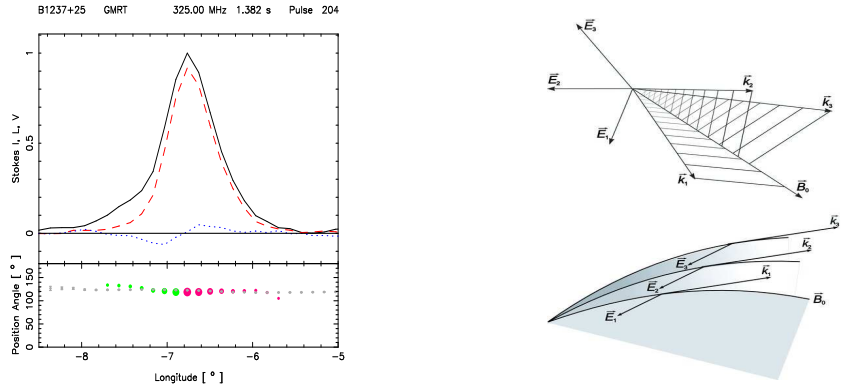}
\caption{The left panel displays a highly polarized subpulse in a single 
pulse of PSR B1237+25. The top panel shows the total intensity (black), 
linear (read dashed) and circular (blue dotted) curves and bottom panel
red (for negative circular) and magneta (for positive 
circular) points show the PPA for the subpulse overlayed on the mean PPA
given by gray points. The right hand plot shows the expected change of the
electric vector i.e. the PPA across the subpulse for two cases, one
for maser emission (upper panel) where the PPA alters quickly across 
the pulse and second for curvature radiation (lower panel) where the 
PPA follows the dipolar magnetic field line planes. These plots are
reproduced from Fig. 1 and 3 of \citet{2009ApJ...696L.141M}.}
\label{fig4}
\end{figure*}
The RVM is a highly successful model for the PPA even when the PPA appear to be
complex and inscrutable.
In a majority of such cases the PPA in the single pulses can be separated into
two orthogonal polarization modes (OPM) with each of the modal tracks following
the RVM, as illustrated in the pulsar B0329+54 (see Fig.\ref{fig2}, third panel
and also \citealt{1995MNRAS.276L..55G,2007MNRAS.379..932M}). 
The RVM requires both the linearly polarized electric and the dipolar magnetic 
field line planes from the beamed radiation to change in the same manner across
the observers line of sight, making it impossible to fix the relative 
orientation of the electric polarization with respect to the magnetic field 
planes using only RVM. 
There are alternate observing schemes used to determine the orientation of the 
emerging polarization direction with the most direct evidence based on 
the X-ray image of the Vela pulsar's wind nebula (PWN). The high resolution 
image from the CHANDRA X-ray telescope (\citealt{2001ApJ...552L.129P, 
2001ApJ...556..380H}) 
shows two symmetric arcs coinciding with the direction of the proper motion, 
$\Psi_{PM} = -53^{\circ}.9 \pm 1^{\circ}$. \citet{2001ApJ...549.1111L} argued 
that the arcs are produced by the relativistic pulsar jets flowing out from the 
neutron star, and the intersection of the arcs coincide with the pulsar 
rotation axis. The PPA traverse of the Vela pulsar has one single track which 
is well modelled by the RVM. \citet{2001ApJ...549.1111L} used the absolute PPA 
measurement from \citet{1999A&A...351..195D} and found the SG point to have a 
value of~$\Psi_{\circ}=35\pm10$. The difference $\mid \Psi_{PM}-\Psi_{\circ} 
\mid = 89^{\circ}\pm11^{\circ}$, suggests that the the emerging polarization is 
perpendicular to the magnetic field line planes. 

The Vela PWN is the only known case where these studies could be conducted. To 
circumvent this, other studies have assumed the direction of the proper motion 
of the pulsar as an indicator of the projection of the rotation axis with the 
Vela PWN serving as a prototype.
In this context \citet{2007MNRAS.379..932M} and \citet{2014ApJ...794..105M}  
investigated PSR B0329+54 and PSR B2045-16 and established that the 
polarization direction for the two RVM tracks are parallel and perpendicular to 
the magnetic field line planes. \citet{2005MNRAS.364.1397J,2007ApJ...664..443R,
2012MNRAS.423.2736N,2013MNRAS.430.2281N,2015MNRAS.453.4485F} estimated a 
distribution of $\mid\Psi_{PM}-\Psi_{\circ}\mid$ for several pulsars and 
established a bimodal distribution around 0$^{\circ}$ and 90$^{\circ}$. 
\citet{2015ApJ...804..112R} found that in core dominated pulsars the 
distribution is mostly around 90$^{\circ}$. 
If we assume the pulsar rotation axis to be along its proper motion, the 
bi-modality in the distribution can be explained by the presence of OPM, with
the emerging radiation either parallel or perpendicular to the magnetic field 
planes. Alternatively, PMs of pulsars can also be parallel or perpendicular
to the rotation axis. While both these explanations are possible, it is evident
that the electric vectors of the electromagnetic waves which detach from the 
pulsar magnetosphere is either parallel or perpendicular to the magnetic field planes.

The implications of the direction of electric polarization are discussed at 
length in \citet{2009ApJ...696L.141M} and \citet{2014ApJ...794..105M}. The 
estimated emission heights $h_{em}$ is the location where the emission escapes 
from the pulsar magnetosphere. It is possible that the emission is generated in
the magnetospheric plasma at a lower height $h_g$ ($< h_{em}$).
In a hypothetical framework one can imagine the two OPM to be generated at 
$h_g$ in an arbitrary orientation with respect to the magnetic field line 
planes. In such a scenario an additional mechanism would be required to rotate
the polarizations either parallel or perpendicular to magnetic field line 
planes as they emerge from the plasma at $h_{em}$.
\citet{1979ApJ...229..348C} suggested that adiabatic walking in the region 
between $h_g$ and $h_{em}$ (see \citealt{1986ApJ...302..120A}) can cause any 
chaotic orientation of the polarization to be rearranged such that they reflect
the orthogonal modes in the RVM. However, this does not automatically orient 
the polarization along parallel or perpendicular to the magnetic plane, which 
would require additional constraints.
The curvature radiation mechanism on the other hand can naturally distinguish 
the magnetic field line planes.
The radiation in the plasma splits into ordinary (O-mode, polarized in the 
plane of the wave vector ${\bf k}$ and the ${\bf B}$ plane) and the 
extraordinary (X-mode, polarized perpendicular to the ${\bf k}$ and ${\bf B}$
plane) waves. 
At emission heights of about 500 km the O-mode can strongly interact with 
the plasma and emerge as a weaker mode or not emerge at all, while the X-mode 
can escape the plasma at $h_g = h_{em}$ as in vacuum
\citep{2014ApJ...794..105M}.
The power in the O-mode is seven times stronger than the X-mode, and hence in 
vacuum the O-mode should be the dominant mode of the radiation. The emergence 
of the X-mode in the Vela pulsar and several other cases provides strong 
evidence that plasma processes are responsible for the coherent radio emission.

The coherent radio emission from pulsars can be excited by a maser mechanism or
via coherent curvature radiation. \citet{2009ApJ...696L.141M} postulated that 
near 100\% linearly polarized single pulses (see Fig.\ref{fig4}, left 
panel) can be used to distinguish between the two emission mechanisms. These 
highly polarized pulses are not depolarized by the OPMs, and hence carry 
information of a single polarization mode. They argued that in the case of 
maser mechanism the {\bf k} vector can be oriented in any direction with 
respect to the local magnetic field (see top panel on the right in 
Fig.\ref{fig4}) and the individual subpulses should show PPA swings across
mean PPA traverse (modeled by the RVM). On the other hand the waves excited by
the coherent curvature radiation are polarized either along the {\bf k} and 
local magnetic field plane (O-mode) or perpendicular to the {\bf k} and 
magnetic field plane (X-mode). The single pulse observations suggest that
pulsar radio emission is excited by coherent curvature radiation which is a
definitive solution to the emission mechanism problem.

\subsection{Evidence for non-dipolar surface magnetic fields}
At heights of 300-500 km from the stellar surface, where the pulsar radio 
emission originates, the magnetic field is largely dipolar in nature. However,
on the neutron star surface the magnetic field should be non-dipolar. The large
curvature of the non-dipolar fields are essential for copious pair plasma 
production which in turn leads to the observed radio emission. 
The strongest evidence of non-dipolar fields on the surface is provided by the 
the longest period (8.5 seconds) pulsar J2144$-$3933 
(\citealt{1999Natur.400..848Y}). \citet{2001ApJ...550..383G} argued 
that under the polar cap models for this long period significant pair 
production requires the 
curvature of the field lines to be $\rho_c \sim 10^5$ cm. This in turn requires
the surface field to be around 10$^{14}$ G, which is about 100 times higher 
than the dipolar magnetic field. \citet{2002A&A...388..235G} proposed a likely 
model for surface non-dipolar field which comprises of a global star centered 
dipole and superposition of local small scale anomalies near the polar cap. 
\citet{1999MNRAS.307..459M} showed that strong non-dipolar field should not 
decay significantly over the lifetime of neutron star and 
\citet{2013MNRAS.435.3262G} proposed that crustal hall drift can create such 
anomalous field structures.  

X-ray observations of older pulsars provide estimates of the surface magnetic 
fields by constraining the size of the polar cap. X-ray emission from neutron 
star is a mixture of thermal and non-thermal components. The thermal emission 
can be further separated into two parts, contributions from the whole surface 
and from the polar cap which acts like a hot spot. In older pulsars the surface 
temperature cools down below 0.1 millon kelvin whereas the polar cap can be
maintained at temperatures larger than a few million kelvin due to bombardment 
of relativistic back-streaming particles in the acceleration regions above the 
polar cap. The black-body emission from older pulsars observed in the soft 
X-rays (0.1 Kev - 10 Kev) can be associated with the hot polar caps. The 
estimates of the temperature $T_{bb}$, and known distance to the pulsar giving 
the X-ray luminosity $L_{bb}$ can be used to calculate the area of the polar 
cap $A_{bb} = L_{bb}/\sigma T_{bb}^{4}$. Comparing the estimated polar cap area
with the dipolar polar cap gives a ratio $b=A_{dp}/A_{bb}$, where $A_{dp}=\pi 
r_p^{2}$ and radius $r_p = \sqrt{2\pi R_s^3/cP}$. Invoking conservation of 
magnetic flux an estimate of the surface magnetic field can be obtained 
$B_s = b B_d$.

The above technique has been applied to a number of pulsars to find $b$ (see 
\citealt{2009ASSL..357...91B}; table 1.4 from \citealt{2013arXiv1304.4203S} for
a list of pulsars) with specific examples of $A_{bb} < A_{dp}$ are 
PSR J0108$-$1431 (\citealt{2009ApJ...691..458P}), PSR J0633+1746 
(\citealt{2005ApJ...625..307K}), PSR B1929+10 (\citealt{2008ApJ...685.1129M}), 
PSR B0943+10 (\citealt{2016ApJ...831...21M}), PSR B1822-09 
(\citealt{2017MNRAS.466.1688H}), PSR B1133+16 (\citealt{2017ApJ...835..178S}). 
In older pulsars $b$ roughly lies in the range $10-500$. 
It should be noted that the estimates of temperature and area are highly 
correlated and should be used with caution as evidence of non-dipolar fields. The results are also affected by models of thermal emission that depend 
on neutron star atmosphere. Several studies \citep{1995ApJ...450..883P,
2004ApJ...616..452Z} show that fits with hydrogen atmospheric models give a 
lower effective temperature by a factor of 2 and around 10-100 times larger 
surface areas. Estimates of actual surface area are also complicated by viewing
geometry as well as general relativistic effects \citep{2002ApJ...566L..85B}.

In summary the basic inputs to the pulsar emission models from observations are
that coherent curvature radio emission is excited in pulsars at a height of 
about 300$-$500 km above the neutron star surface in regions of open dipolar 
magnetic field lines and the magnetic field on the neutron star surface are 
significantly non-dipolar in nature.

\section{Plasma condition in the magnetosphere and Mechanism for Radio Emission in Pulsars}
\label{sec3}
The observational results discussed so far are vital inputs into RS75 class of 
models from the polar cap where the coherent curvature radiation is generated 
from charged bunches. In this section we will first discuss the basic 
hypothesis of the RS75 model and the plasma conditions in the magnetosphere. 
Subsequently, we will discuss the observations which are interpreted based on 
the polar-cap model.

\subsection{Gap Formation:} 
RS75 suggested that in pulsars where ${\mathbf \Omega.B_s} < 0$ above the 
magnetic poles, the polar cap is positively charged. Initially there is only a 
limited supply of positive charges above the polar cap which is 
relativistically flowing away along the open magnetic field lines as a pulsar 
wind. If the binding energy of the ions on the neutron star surface are 
sufficiently large, the region above the polar cap will be charge deficient 
thereby creating a vacuum gap with large electric fields. They suggested that 
if a gap of height $h$ exists above the polar cap, the potential drop 
$\Delta V$ across the gap increases as $h^2$ since $\Delta V =  \Omega B_s 
h^2/c$. Such a gap region can discharge by the generation of electro-positron 
pairs by photons of energy $> 2 m c^2$. Considering the diffuse background to 
be the source of such photons the discharge can happen within 100 $\mu$s 
(\citealt{1982ApJ...258..121S}).
These charges further accelerate with Lorentz factors $\gamma$ in curved
magnetic field lines of radius of curvature $\rho$ to produce high energy curvature radiation photon with frequency $(3/2) \gamma^{3}c/\rho$ 
which after traveling a mean free path $l_e$ can produce another
electron-positron pair, such that the gap height $h \sim l_e$.
To find $h$, RS75 used the \citet{1966RvMP...38..626E} condition where pair creation 
can happen if the parameter
$\chi = (\hbar \omega/2m_e c^2)B_s \sin\theta/B_q \approx 1/15$.
Here the critical magnetic field $B_q = m_e^2c^3/e\hbar^2=4.4 \times 10^{13}$ G and
$\theta$ is the angle between the photon and highly curved $B_s$ such 
that $B_s \sin\theta \sim B_s h/\rho$. In terms of pulsar parameters 
writing the dipolar component of the field as $B_d/10^{12} \sim  \sqrt{P \dot{P}_{15}}$ 
where period derivation $\dot{P_{-15}} = \dot{P}/10^{-15}$s/s and $\rho_6 = \rho/10^{6}$ cm,
$h$ and the maximum potential drop across the gap $\Delta V$ 
can be expressed as (along with typical values of $b=10$, $P=\rho_6=\dot{P}_{15}=1$)

\begin{equation}
h = 5 \times 10^3 b^{-4/7} P^{1/7} \dot{P}_{-15}^{-2/7} \rho_6^{2/7} \sim 1350 \rm{cm}
\label{height}
\end{equation}

\begin{equation}
\begin{aligned}
\Delta V = 5.2 \times 10^{9} b^{-1/7} P^{-3/14} \dot{P}_{-15}^{-1/14} \rho_6^{4/7}\\
         \sim 3.7 \times 10^{9} \rm{[statvolt]}
\end{aligned}
\label{deltaV}
\end{equation}

\subsection{Spark Formation:} A number of such localized discharge
can form in the gap and each such discharge undergoes
a pair creation cascade. The electric field in the gap accelerates
the electrons towards
the stellar surface, while the positrons, often called the
primary particles are accelerated away from the surface.
At the top of the gap the primary particles acquires Lorentz factors of $\gamma_p$
such that

\begin{equation}
\gamma_p \approx e \Delta V/m_e c^2 \approx 2 \times 10^6 
\label{gammap}
\end{equation}

As the primary particles move away from the gap region where $\mathbf{E.B} = 0$,
they continue to create high energy photons, which further create pairs
and this cascade leads to the generation of
of a cloud of secondary electron-positron plasma which has a significantly
lower Lorentz factor with mean value of $\gamma_s$.
If there are $n_p$ primary pairs then the number of secondary
pairs can be estimated as $n_s \sim (2\gamma_p/\gamma_s) n_p$ and thus
the density of the secondary plasma increases by a multiplicity
factor $\kappa = n_s/n_p$. The burst of pair-production process
increases the charge density along the gap discharge stream and screens
the potential in the gap. This process happens exponentially
and after a certain time $\tau$ which is estimated to be
$\tau \sim 30-40 h/c \sim 1 \mu$s (RS75), the charge density becomes close to $n_{GJ}$,
and the pair-production process stops.  During this time
the discharge spreads in the lateral direction thus giving a width of $\sim h$
to the discharge. This fully formed discharge is called a ``spark'' and each
spark is associated with a secondary plasma cloud.
Once the spark is formed, the spark associated plasma column leaves the gap
in a time $\tau_g \sim h/c \sim 0.005\mu$s,  and as soon as a height $h$ is reached,
the next discharge can initiate. Thus in the plasma frame of reference
the plasma cloud at a height $h_{em}$ can be thought to be like a cylindrical column
of length 30-40$h \sim 1$km 
along the magnetic axis and having a transverse scale of $h (h_{em}/R_s)^{0.5}\sim 0.1$ km.
lines. The sparking process described above is highly simplistic although
numerical simulation of the pair cascade process in one dimension
by \citet{2010MNRAS.408.2092T} appears
to confirm the non-stationary sparking process.
Detailed 3D simulations are required to get a fully consistant theory of
spark formation. 

\subsection{ Characteristics of the secondary plasma:}
The number density of the secondary plasma cloud is $n_s = \kappa n_{GJ}$ since
$n_p = n_{GJ}$, where $n_{GJ}$ can be expressed in terms of pulsar parameters as 

\begin{equation}
n_{GJ} \sim 6 \times 10^{10} \left(\frac{\dot{P}_{-15}}{P}\right)^{0.5} 
\left(\frac{R_s}{h_{em}}\right)^{3} \sim~~ 5.5 \times 10^5 \rm{cm^{-3}} 
\label{ngj}
\end{equation}
with the numerical value obtained for $R_s =10$ km and $h_{em}=500$km.
Considering radio emission arising at 1 GHz an estimate of $\gamma_s \sim (10^{9}\rho100/c)^{1/3}$ 
is obtained if one considers curvature
radiation from dipolar field lines at $h_{em}=500$ km where $\rho_6 \sim 100$ cm.
This gives $\kappa \sim (n_p/n_s) \sim \gamma_p/\gamma_s\sim 10^4$ (e.g. RS75).
The corresponding plasma frequency of the secondary plasma $\omega_p^2=4\pi e^2 n_s/m_e$ is given by

\begin{equation}
\omega_p = 4.3 \times 10^{11} \kappa^{0.5} \left(\frac{\dot{P_{-15}}}{P}\right)^{0.25}
\left(\frac{1}{h_{em}}\right)^{1.5}~\sim~~3.7\times10^9 \rm{Hz} 
\label{omegap}
\end{equation}

The properties of the secondary plasma provides interesting
constraints for the radio emission as was discussed by 
\citet{2014ApJ...794..105M} and below we review their arguments.
The plasma properties of the secondary plasma can be 
written in the observer frame of reference in terms of 
the basic pulsar parameters 
$P$ (in sec) and ${\dot P}$, and as a function of the 
fractional altitude $\Re=r/R_{\rm LC}$ (here
$r$ is the radial distance from the center of the neutron star and $R_{\rm LC}=4.8\times 10^9P$ cm is the radius of the light cylinder). 
The cyclotron frequency $\nu_B$ in the local magnetic
field $\mathbf{B}$ can be expressed in GHz as,

\begin{equation}
\nu_B = \frac{\omega_B}{2\pi\gamma_s} = 5.2 \times 10^{-2} \frac{1}{\gamma_s} \left(\frac{{\dot
P_{-15}}}{P^5}\right)^{0.5} \Re^{-3}. 
\label{nu_B}
\end{equation}

The characteristic  plasma frequency $\nu_{\circ}$ in GHz can be written as:
\begin{equation}
\nu_{\circ} = \frac{\omega_0}{2\pi}= 2\times 10^{-5}\kappa^{0.5} \sqrt{\gamma_s}\left(\frac{{\dot
P_{-15}}}{P^{7}}\right)^{0.25}\Re^{-1.5},
\label{nu_0}
\end{equation}
where in the observer frame the plasma frequency is $\omega_{\circ}=2\sqrt{\gamma_s}\omega_p$.
As the third parameter we consider the
characteristic frequency of the charged solitons (discussed later) coherent curvature radiation, 
which correspond to the maximum of the power spectrum $\nu_{\rm cr}$ 
(see Fig. 2 in \citealt{2004ApJ...600..872G}, GLM04 hereafter),

\begin{equation}
\nu_{\rm cr} = 1.2 \frac{c {\Gamma}^3}{2 \pi \rho}= 0.8 \times 10^{-9}
\frac{{\Gamma}^3}{P}\Re^{-0.5}, 
\label{nu_cr}
\end{equation}
$\Gamma$ is the Lorentz factor of the solitons which is slightly different from $\gamma_s$ (see 
GLM04).

The right panel of Fig.~\ref{fig3} (reproduced from \citet{2014ApJ...794..105M})shows
$\nu_B$ (red line), $\nu_{\circ}$ (green line) and $\nu_{cr}$ (blue line) 
for a pulsar with $P$= 1 sec and
${\dot P}_{-15}$=1, in the observers frame of reference as a function of $\Re$.
For the plots a typical value of $\gamma_s \sim 200$ is used, and since 
$\kappa$ has some uncertainty, two extreme value of $\kappa = 10^2$ and $10^4$ has
been chosen while plotting $\nu_{\circ}$. Similarly the $\nu_{cr}$ curves are 
plotted for two extreme values of $\Gamma = 300$ and 600.  
The horizontal gray line correspond to 
the frequency range of observable radio emission (about 10MHz-10GHz) range 
and what is noteworthy is that typically $\nu_{cr} < \nu_{\circ} << \nu_{B}$ 
for $\Re < 0.1$ which is where the radio emission originates (see left panel Fig.(\ref{fig3}). 
The shaded region correspond to the region where two stream instability
can grow, which we will discuss in the next section.

\subsection{ Subpulse drift \& $E \times B$ of sparks:}
The phenomenon of subpulse drifting discovered by 
\citet{1968Natur.220..231D} is undoubtedly fascinating. In some 
pulsars, when single pulses 
are displayed as a pulse stack, where consecutive pulse periods are arranged on 
top of each other in a two dimensional array with the abscissa as pulse phase and 
ordinate as pulse longitude, the subpulses in the pulse window are seen 
to systematically move or drift in pulse phase from one edge to the other edge of 
the pulse window. Thus an impression of drift bands is seen 
in the pulse stack where the drifting pattern repeats itself with a frequency $f_3$ (or time $P_3 = 1/f_3$). The usual technique 
employed (called longitude resolved fluctuation spectra (LRFS), see 
\citealt{2001MNRAS.322..438D}) to recover $f_3$ is by 
performing fast Fourier transforms for every pulse phase within the 
pulse window along the ordinate axis. Drifting greatly varies in 
the pulsar population and based on the variation of the phase of the Fourier 
transform is broadly divided into two groups: phase modulated 
drifting and amplitude modulated drifting. In the catagory of phase modulated 
drifting pulsars, where the phase has a positive slope from the leading to the trailing
edge, are called negative drifting (ND) while the negative slope cases are 
called positive drifting (PD). The pulsars where the phase is constant across
longitude and only the intensity is modulated, are called amplitude
modulated drifting (AMD). Detailed single pulse studies has revealed that 
the drifting phenomenon in the form of ND, PD or AMD exist for 
about 45\% of the pulsar population (e.g. \citealt{2006A&A...445..243W}, \citealt{2016ApJ...833...29B}).

Drifting is intimately connected with the dynamics of the radio emitting
plasma and the sparking model of RS75 is the only successful model that is currently
invoked to explain this phenomenon. In the RS75 model, the observed subpulses 
emitted over a bundle of magnetic open field lines in the dipolar 
radio emission region ($h_{em}$) 
can be traced down to the spark associated plasma column in  non-dipolar polar cap.  
The IVG in RS75 model initially has a 
co-rotational electric field $\mathbf{E_{cor} = -(\Omega \times r)\times B}/c$, and then 
the gap discharges in the form of multiple sparks. As each spark 
develops, it initially lags behind co-rotation, and eventually starts to
co-rotate when
the fully formed spark attains $n_{GJ}$ and the electric field gets entirely
screened. During the sparking process, the spark pattern lags behind 
the co-rotation, although the foot of the sparking location drifts slightly, and hence
the new spark initiates at this displaced location. This gives rise to the 
drifting pattern, and the velocity of the drift motion being 
$\mathbf{v_{cor} = (E_{cor}\times B)}c/\mathbf{B}^2$. RS75 considered an 
anti-pulsar where the rotation and magnetic axis are aligned but opposite to each 
other ($\mathbf{\Omega \cdot B = -1}$), and a number of 
sparks $n_{sp}$ perform drift motion both around the 
rotation and magnetic axis. Assuming that the sparks perform a circular motion 
over a time $\hat{P_3} = P_3 n_{sp}$ around the polar cap, RS75 found the expression 
$\hat{P_3} \approx 5.6 B_{12}/P^2$ in units of $P$.  
$\hat{P_3}$ cannot be easily inferred from observations, but 
in a detailed analysis of the drifting pulsar 
PSR B0943+10 \citet{2001MNRAS.322..438D} found evidence of a tertiary 
periodicity in their
LRFS analysis, which they interpreted was due to the spark circulation 
time $\hat{P_3} = 37 P$, whereas the spark repeating time $P_3 = 1.8 P$, 
gave $n_{s} = 20$. To explain this result they proposed the
carousel model (which is the anti-pulsar drift model of RS75) for 
PSR B0943+10 where 20 sparks are circulating around the pulsar magnetic 
axis and 
$\hat{P_3}$ is the time taken for one spark to return back to the same location.
Theoretically the IVG model of RS75 predicts a much 
smaller value of $\hat{P_3} \approx 11 P$ for PSR B0943+10, and this led Gil et al. (2004) 
to introduce partially screened 
gap (PSG) model where the gap potential is screened by an amount $\eta$ due 
to presence of ions. The carousel model has also been invoked to explain
the nested core-cone structure in pulsar emission. Gil \& Sendyk (2000) 
proposed that the polar cap has a set of circular sparks
where the sparks touch each other and the radius of the spark
is equal to the gap height $h_g$. The central spark remains stationary in phase
and correspond to the core emission, while the other sparks rotate around
the central spark (like a carousel) to produce the conal emission.

\begin{figure}[t!]
\includegraphics[height=.23\textheight,angle=0]{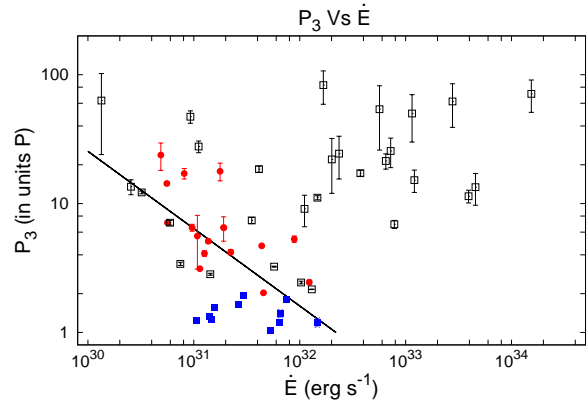}
\caption{The figure shows the variation of drifting periodicity 
$P_3$ with spin-down energy $\dot{E}$. The red points correspond to PD 
case, the blue points correspond to ND case and the rest correspond to AMD case. The black line correspond to eq.(\ref{eq16}). The figure is reproduced
from Fig. (6) of \citet{2016ApJ...833...29B}.}
\label{fig5}
\end{figure}

\citet{2016ApJ...833...29B} recently
pointed out a fundamental difficulty in connecting
the carousel model to pulsar data since in a real pulsar 
the magnetic axis and the rotation axis are not aligned. 
In such a case the direction of the sparks motion that lag the
co-rotation velocity is around the rotation axis and not the magnetic 
axis. An external observer
hence would essentially see that the sparks are drifting across the
polar cap. In fact the carousel model for 
non-aligned pulsar will contradict the basic assumption of the RS75
model as parts of the spark motion in the polar cap will lag the co-rotation
velocity while there will be parts that will be leading. For PSR B0943+10 
(discussed earlier) invoking a carousel is not essential to interpret $\hat{P_3}$. 
The viewing geometry for this pulsar is highly tangential and the sparks essentially
moves along the line of sight, which makes it difficult to infer the path along which the spark
moves in the unseen part of the pulsar beam. In the absence of a carousel the periodicity $\hat{P_3}$ 
might have a different physical origin.

In a careful analysis of the MSPES dataset \citet{2016ApJ...833...29B} applied the concept of lagging of sparks
to ND and PD pulsars. They argued that $P_3=1/f_3$ corresponds to the ND cases 
where sparks lags behind co-rotation, while for PD cases the phase slope appear to be opposite 
due to aliasing effect and hence $P_3=1/(1-f_3)$. Once this is taken into account
a remarkable anti-correlation is seen between $P_3$ and the slowdown energy 
$\dot{E}$ of the pulsar and can be expressed as 
\begin{equation}
P_3 = (\dot{E}/2\times 10^{32})^{-0.6} 
\label{eq16}
\end{equation}
This dependence is seen in Fig.(\ref{fig5}) where the black line correspond
to eq(\ref{eq16}).
Also note that phase modulation drifting is only seen below 
$\dot{E} < 2\times 10^{32}$ ergs/s, and above this value only phase stationary 
amplitude modulation is observed. It then appears that the drifting (PD and ND) 
phenomenon in pulsars can only be observed for $\dot{E} < 2\times 10^{32}$ cases.
Beyond this value $P_3 < P$ and hence $P_3$ cannot be measured. Thus the AMD cases for
$\dot{E} > 2\times 10^{32}$ ergs/s might be a entirely new phenomenon as is argued
by \citealt{2016ApJ...833...29B} and \citet{2017arXiv170309966M}. Under the PSG model the 
actual drift velocity of sparks is $v_d = \eta v_{cor}$, and if one assumes
that the polar cap is filled with sparks of size $h$ such that the distance between
the sparks is $2h$, then the spark repeating time $P_3 = 2h/v_d$. It can be shown
the if $\eta$ is small ($\sim 0.1$) then $P_3 = 1/(2\pi \cos(\alpha) \eta)$ (see eq(3.55) \citealt{2013arXiv1304.4203S}) 
where $\alpha$ is the angle between the rotation and magnetic axis. The factor 
$\eta \cos(\alpha) \propto  \dot{E}^{0.5}$ and hence a dependence
of $P_3 \propto E^{-0.5}$ as see in eq(\ref{eq16}) can be obtained (see eq. (12) in \citet{2016ApJ...833...29B}. 
Note that the sparks lags behind co-rotation at the polar cap where the magnetic field is
non-dipolar however the spark associated plasma cloud in the emission region is dipolar.
Hence the lagging motion of the spark in the polar cap w.r.t the observers line of sight 
can be very different when it translates to the observed subpulse motion 
in the emission region at $h_{em}$.

A large number of radio pulsar phenomenon such as pulse microstructure (not
discussed here, see e.g. \citealt{2015ApJ...806..236M} for a recent study), shape of pulsar beam, and subpulse
drifting, are thought to be related to the sparking process in the IVG.
it is important to state that the physics of how the sparks develops in the polar
cap is still very unclear, and more insights are necessary to connect theory 
with observations. 
\begin{figure}[!t]
\includegraphics[height=.33\textheight,angle=0]{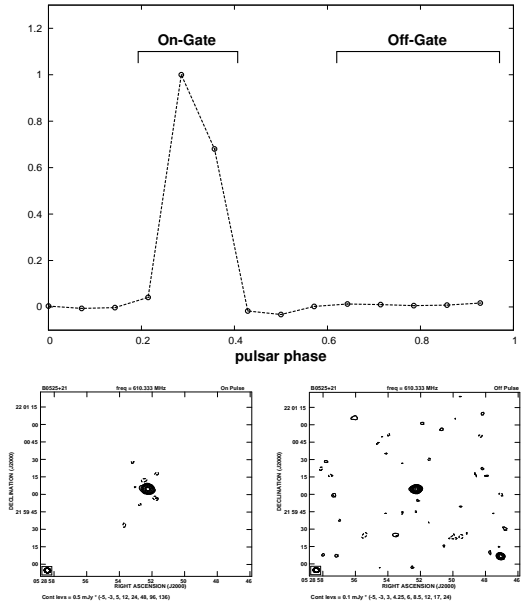}
\caption{The top panels show the average pulse profile of PSR B0525+21 
obtained by folding the interferometric time series data 
from the Giant Meter-wave Radio Telescope (GMRT) at 325 MHz. 
The pulse phase regions for the on and off pulse gates are indicated. 
In the lower panel the contour maps of the on-pulse (bottom left, the
on-pulse flux is 21.1$\pm$1.5 mJy)  
and off-pulse (bottom right, the off-pulse flux is
3.6$\pm$0.3 mJy) emission from the pulsar. The figure is reproduced from 
Fig. (1) of \citet{2012ApJ...758...91B}.}
\label{fig6}
\end{figure}

\subsection{Off-Pulse Emission, Cyclotron resonance of primary and secondary plasma at outer magnetosphere} 
In pulsar time series analysis, the off-pulse (OP) region is noise-like without
any structure. In certain studies, the search for
continuum OP emission has been 
done by using the technique of gated interferometry. 
These studies are mostly done to search for radio emission
from pulsar wind 
nebula (PWN) associated with young and energetic pulsars, 
where the outflowing particles from the pulsar wind 
can interact with the interstellar medium to produce
synchrotron emission observable in the radio band (see e.g. 
\citealt{2000MNRAS.318...58G}). Generally for older pulsars, the 
energy in the pulsar wind is not sufficient to power a PWN, and 
hence one does not expect to see any PWN related emission
in the OP. A confirmation of this came from 
the work of \citet{1985MNRAS.212..489P} where no definite OP emission 
above 1 mJy was found after performing the search in several older pulsars.   

Recently \citet{2011ApJ...728..157B,2012ApJ...758...91B} conducted a study of OP emission 
in a number of long period and low energetic pulsars 
and after doing a very careful analysisi, they surprisingly 
discovered OP emission at the flux level of a few 
mJy (see Fig.\ref{fig6}). These pulsars however
have low slowdown energy hence any emission from the PWN
can be ruled out. Further, they demonstrated
that the OP emission scintillates in the same manner as
the MP and this confirmed that the OP emission has a 
magnetospheric origin and is coherent in nature.
\citet{2013ApJ...772...86B} proposed that the OP emission might
be originating in regions close to the light cylinder where 
the coherent radio OP emission might arise due to 
growth of cyclotron resonance instability in the outflowing magnetospheric
plasma (\citealt{1991MNRAS.253..377K}, \citealt{2000MNRAS.315...31L}). 
The mechanism 
can be understood as follows. Since 
the magnetic field near the light cylinder is weak (a few 
gauss), the outflowing 
secondary pair plasma (generated close to the IVG in the polar-cap 
models) can gyrate around the magnetic field. A large number
of electromagnetic modes exists in the plasma, and the transverse X-mode
is one such mode which is capable of propagating in the plasma and escape
as electromagnetic wave. As the wave propagates, amplification to
this wave can be provided by exchange of energy between 
the wave and fast moving primary plasma particles via the well
known cyclotron resonance instability. 
\citet{2013ApJ...772...86B} derived the growth rate of this instability
and determined the magnetospheric OP emission are likely to 
be detected in pulsars where the dimensionless growth rate $\Gamma \tau > 1$, where  
\begin{equation}
\Gamma \tau = 9.3 \left(\frac{P^3}{\dot{P}_{15}}\right)
\label{eq17}
\end{equation}
(see eq.(16) of \citealt{2013ApJ...772...86B}).
Note that the MP emission arises from growth of instability in 
secondary plasma while the OP emission 
requires interaction between both primary and secondary pair-plasma. 
Thus, OP emission can be considered as the most direct evidence for
the generation of the two component pair-plasma in the pulsar magnetosphere.
Application of eq.(\ref{eq17}) is currently limited due to lack of
sensitive instruments that can detect the low level OP.

\section{\bf Growth of Linear Two Stream Instability in Secondary Plasma}
In the previous sections we have gathered all evidences that support
the polar-cap RS75 kind of model in pulsars. In this section we will
discuss the mechanism that excite coherent curvature radiation in pulsars.
Recall that in the IVG the non-stationary sparking process can 
generate flow of successive plasma cloud 
strictly flowing along a bundle of magnetic dipolar field lines.
The non-stationary flow gives rise to situations where slow 
and fast moving particles can overlap and hence the crucial
two-stream instability can develop in the pulsar plasma 
(RS75, \citealt{1977MNRAS.179..189B}, 
\citealt{1983Ap.....19..426E}, \citealt{2002nsps.conf..240U}
for a review). 
Each plasma cloud has a mean Lorentz factor $\gamma_s$  
and there is a sufficient wide spread in $\gamma_s$ which arises due to the 
pair cascade process with typical minimum $\gamma_{min} \sim 10$ and
maximum values of $\gamma_{max} \sim 10^3-10^4$.
The overlapping to the slow and fast moving particles of two successive clouds can lead to
the two stream instability in plasma. Usov (1987) used simple kinematical
estimates to show that this instability can be important for pulsar
emission mechanism and that it can develop in the 
pulsar emission region. Following \citet[hereafter MGP00]{2000ApJ...544.1081M}, 
the typical velocity difference between 
between the slow and fast moving particles is about $\Delta v=c/(2\gamma_s^2)$, 
and the typical time for the particles to overlap is $\Delta T= h/\Delta v \sim 
2\gamma_s^2 h /c$. Hence, the instability can develop at a distance
$r\sim c\Delta T \sim 2 \gamma_s^2 h$ which can be written in terms
of the pulsar parameter as,
\begin{equation}
r/R \sim 10 \left(\frac{\gamma_s}{100}\right)^2 \rho_6^{2/7} B_{12}^{-4/7} P^{3/7}
\label{rin}
\end{equation}
For typical values of parameters $r$ is about a few hundred km which 
agrees well with the observed emission heights $h_{em}$ derived for pulsars
(The instability region is indicated as the shaded region 
in right panel of Fig.(\ref{fig3})).
A significantly detailed study of the two-stream instability considering
overlapping of multiple clouds as in the real case was considered
by \citet{1998MNRAS.301...59A} and they found instabilities can grow if
\begin{equation}
1.1 \times 10^{4}(\gamma_s/100)^{-1.5} r^{-1.5} (\dot{P}_{-15}/P)^{0.25} >> 0.1 
\label{eq18}
\end{equation}
(see eq.(7) of MGP00). This condition can easily
develop in the secondary plasma for $\gamma_s \sim 100$ and hence the two-stream instability
can excite strong electrostatic unstable Langmuir waves, with frequency
$\omega_l$ which in the observer frame is given by.
\begin{equation}
\begin{aligned}
\omega_l = 2 \delta_w {\gamma_s} \omega_p \approx 
 4.3 \times 10^{11} \kappa^{0.5}\gamma_s \left(\frac{\dot{P_{-15}}}{P}\right)^{0.25}
\left(\frac{1}{h_{em}}\right)^{0.25}\\
~~\sim~~ 37 \times 10^{10} \rm{Hz}
\end{aligned}
\label{omegal}
\end{equation}
where the parameter $\delta_w\sim 0.5$ (see \citet{1998MNRAS.301...59A})

\section{Coherent Radio emission: Emission From Bunches}
Langmuir waves are electrostatic waves, and as they develop in the plasma
they tend to bunch the charges with typical length
of half the Langmuir wavelength. RS75 and \citet{1979ApJ...229..348C} suggested 
that charged bunches formed by linear Langmuir
waves can emit coherent curvature radiation near the plasma frequency.
This is however impossible as was shown by (\citealt{1986FizPl..12.1233L}, 
MGP00 2000, \citealt{2014ApJ...794..105M}).
They argued that curvature radiation with frequency close to the
local plasma frequency is impossible, because 
the coherence condition requires the characteristic dimension of the bunches to 
be shorter than the wavelength of the radiated wave, which can never be met as the
bunch would disperse before the radiation is emitted. 
Hence high-frequency Langmuir plasma wave cannot be responsible for the coherent
pulsar radio-emission (see also \citealt{1999ApJ...521..351M}).
The other constraint is from observations where the frequency of radio 
emission is significantly lower than $\omega_l$.

Alternatively MGP00 proposed a model for pulsar
radio emission based on the modulational instability of Langmuir
waves, the basic outcome of the theory is discussed here. MGP00 argued that
due to the thermal spread in the plasma
the frequency of the Langmuir waves are likely to have a small
spread $\Delta \omega$, such that $\Delta \omega << \omega_l$.
The amplitude of the Langmuir wave packet will hence be modulated
by low-frequency beatings with phase velocity $\Delta \omega /\Delta k$
which is also equal to the group velocity of the plasma wave $v_g = d\omega/dk$.
Noting that the plasma wave moves in the electron-positron plasma
which is also moving relativistically, resonant interactions of plasma
particles with low-frequency beatings will lead to modulational instability.
Such a process is described by the non-linear Schr\"{o}dinger equation (NLSE)
with non-linear Landau damping. \citet{1973JPSJ...34..513I} derived the
NLSE taking into account the effect of Landau damping for non relativistic
case and applied it to an electron ion plasma. \citet{1980Ap&SS..68...61P}
derived the NLSE for the relativistic case and MGP00 applied it to the pulsar system in detail.

Considering the landau damping term to be small they 
found the corresponding soliton solution by solving the NLSE. 
In general the Lorentz factors of electrons and positrons in the secondary 
plasma has slightly different distribution function ($\Delta \gamma = \mid \gamma_{+}
-\gamma_{-}\mid$) due to 
$\Omega.B$ changing along the magnetic field lines (\citealt{1977ApJ...212..800C}
, 
\citealt{1998MNRAS.301...59A}). If this difference $\Delta \gamma$ is large, then a charge separation
can be obtained within the envelope of the soliton (i.e. charged bunches). MPG00 found for reasonable pulsar
parameters $\Delta\gamma/\gamma \sim 1$ which was adequate for the various assumptions
in the theory like growth of linear instability and derivation of the NLSE to work.
The charged solitons thus obtained are supported by pondermotive force (Gaponov
\& Miller 1958) and are capable of generating coherent curvature radiation. 
MGP00 found the soliton size $\Delta_s \sim 10-100$ cm
and coherency of curvature radiation process 
wavelength of the emitted waves should be longer than the longitudinal size of the soliton $\Delta_s$.
Thus, the frequencies plotted in Fig. (\ref{fig3}) should obey the following constraints
$\nu_{cr} < c/\Delta_s < \nu_{circ} << \nu_{B}$
which is clearly the case. The observed pulsar radiation cannot be generated at
altitudes exceeding 10\% of the light cylinder radius which is consistent with observations. 
MGP00 showed that in a plasma cloud about 10$^5$ solitons can form with each soliton 
have charge of $Q=e10^{21}$. The power of curvature radiation by a charge $Q$ is 
$P \approx Q^2 c \gamma_s^4/\rho^2 \sim 10^{25}$ ergs/s, and for a soliton case
MGP00 showed that it is 100 times smaller. The incoherent addition of power form the whole
soliton cloud multiplied by several number of sparks (say 10) gives the emitter power
to be $\sim$10$^{29}$ ergs/s which can account for the observed radio luminosity in pulsars. 

The curvature radiation by solitons excite the transverse X and O mode below
$\omega_{\circ}$ which travels with two different speed and
the refractive index is such that the X-mode can 
emerge as in vacuum while the O-mode gets ducted away along the field lines.  
\citet{2014ApJ...794..105M} showed that the difference between the refractive 
index is constant and hence there is no adiabatic walking of pulsar radiation.
There are however two observational effects that still needs to be understood. One 
effect is the presence of OPM which requires the O-mode to emerge
from the plasma, and one suggestion is that gradient in plasma density 
can cause the O mode to emerge. The second effect is that 
large amounts of circular polarization are observed, however if
the X and O mode separates in phase, 
the circular polarization is lost, and hence a
new mechanism is needed to expalin the circular polarization. Finally, MGP00
ignored the effect of landau-damping while solving the NLSE and hence got
stable time-independent soliton solutions. The stability of these solitons
needs more investigations. 
 
\section{Concluding Remarks}
The coherent radio emission from pulsars are intricately linked to the pair 
creation process around the star. The different plasma parameters are 
controlled by the pair creation process and sensitive to any stochastic 
variations.
The minimum timescales for any change in the emission process is around 100 
milliseconds for a typical pulsar, which is the time taken by the return 
current to traverse the pulsar magnetosphere.
On the other hand a change in the emission state for longer timescales requires
additional source of pair creation to alter the nature of the inner 
acceleration region at longer timescales.
The phenomenon of mode changing and nulling in pulsars 
(not discussed here; see e.g. \citealt{2017ApJ...846..109B}) 
where a sudden change in the emission state is observed and persists for longer
timescales of varying length is indicative of such changes.
Understanding these phenomena requires new physical insights which are outside
the purview of the steady state pulsar emission model discussed in this paper.


\section*{Acknowledgement}

The author would like to acknowledge his colleagues
and collaborators with whom he had been engaged
in pulsar research, in particular, Janusz Gil, Joanna
Rankin, G. Melikidze and Rahul Basu. He thanks Rahul
Basu for discussions, critical reading and giving constructive
comments on this manuscript. He also thanks
the organizers for giving him the opportunity to write
this review in honour of Prof. G. Srinivasan who has
been his mentor and a source of inspiration.

\bibliography{mnrasmnemonic,gs75_mitra_jaa}
\bibliographystyle{mnras}
\end{document}